\documentclass[12pt]{iopart}

\usepackage[dvips]{graphicx}

\newcommand{\gdsn}{Gd$_2$Sn$_2$O$_7$}
\newcommand{\gdti}{Gd$_2$Ti$_2$O$_7$}
\newcommand{\ygdsn}{(Y$_{0.995}$Gd$_{0.005}$)$_2$Sn$_2$O$_7$}

\newcommand{\Stev}[2]{\hat{O}^{#1}_{#2}}
\newcommand{\BStev}[2]{B^{#1}_{#2}\hat{O}^{#1}_{#2}}
\newcommand{\pyr}[2]{#1$_2$#2$_2$O$_7$}

\begin{document}

\title[Single-ion anisotropy in  \gdsn{}...]{ESR study of the single-ion anisotropy in the
pyrochlore antiferromagnet \gdsn{}.}

\vspace*{5mm}

\author{ V.N.Glazkov$^{1,2}$, A.I.Smirnov$^1$, J.P.Sanchez$^2$, A.
Forget$^3$, D. Colson$^3$, P. Bonville$^3$}

\address{$^1$ P.L.Kapitza Institute for Physical Problems, RAS, 119334
Moscow, Russia}

\ead{glazkov@kapitza.ras.ru}

\address{$^2$ CEA, Centre d'Etudes de Grenoble,
DRFMC/SPSMS, 38054 Grenoble, France}

\address{$^3$ CEA, Centre
d'Etudes de Saclay, DRECAM/SPEC, 91191 Gif-sur-Yvette, France}

\begin{abstract} Single-ion anisotropy is of
importance for the magnetic ordering of the frustrated pyrochlore
antiferromagnets \gdti{} and \gdsn{}. The anisotropy parameters
for the \gdsn{} were measured using the electron spin resonance
(ESR) technique.  The anisotropy was found to be  of the easy
plane type, with the main constant $D$=140\,mK. This value is 35\%
smaller than the value of the corresponding anisotropy constant in
the related compound \gdti{}.
\end{abstract}


\submitto{\JPCM}

\section{Introduction.}

 Rare earth pyrochlore magnets \pyr{R}{M}{}
 (R --- rare earth ion, M --- transition metal) have attracted recently
 a lot of attention due to the specific geometry of exchange  bonds.
 Rare earth ions in a pyrochlore structure form a network of corner-sharing
 tetrahedra.
The nearest-neighbors antiferromagnetic (AF) exchange interaction
is strongly frustrated in this lattice (see, e.g.,
\cite{ramirez-rev}) and the classical ground state of this system
should remain macroscopically degenerate down to T=0.

Selection of a unique ground state in the real magnets should
occur due to other interactions like further neighbor exchange
interactions, dipole interactions, single-ion anisotropy or due to
the lifting of degeneracy by fluctuations. Single-ion anisotropy
is known to have a strong effect on the formation of the ground
state: strong axial anisotropy favors an unusual spin-ice state in
\pyr{Dy}{Ti}{} \cite{higashinalka} while in the case of a strong
easy-plane anisotropy in \pyr{Er}{Ti}{}  a N\'eel state is formed
by the quantum order-by-disorder mechanism \cite{champion}.

Gd-based pyrochlore magnets \gdti{} and \gdsn{} were considered to
be  real examples  of an AF Heisenberg pyrochlore. Both compounds
demonstrate a magnetic ordering at a temperature near 1\,K, but
their spin structures are quite different. In the case of \gdti{},
two successive transitions  are observed at 1.02\,K and 0.74\,K
\cite{bonville}. The low-temperature magnetic structure is a
complicated noncollinear multiple-$\bi{k}$ structure
\cite{gdti-structure}. In the case of  \gdsn{}, a unique phase
transition is observed \cite{bonville} and the magnetic structure
is a non-collinear with $\bi{k}=0$ \cite{gdsn-structure}.

Up to now, single-ion anisotropy effects were not considered for
these compounds  since the Gd$^{3+}$ ion is an S-state ion with
$L$=0 \cite{ramirez-prl}. However, a considerable anisotropy was
found in \gdti{} \cite{glazkov}. A crystal field splitting of the
ionic levels occurs because of a strong spin-orbit coupling among
$4f$-electrons, which breaks the simple $LS$-scheme of the energy
levels and leads to the admixture of the $L\neq0$ states into the
ground state of the Gd$^{3+}$ ion. The main anisotropy constant
$B^0_2$  was found to be equal to 74\,mK in \gdti{}, which makes
the overall crystal field splitting $3B^0_2S^2 \simeq$2.7\,K
comparable with the exchange energy $JS^2\simeq$3.7\,K. Thus, it
is important to determine the values of the single-ion anisotropy
constants in \gdsn{} as well. A difference in the single-ion
anisotropy could be important for understanding the above
mentioned difference of the magnetic ground states in \gdti{} and
\gdsn{}.

To study single-ion effects, we used the non-magnetic
isostructural compound \pyr{Y}{Sn}{}  with a small amount
(nominally 0.5\%) of gadolinium substituting for yttrium. This
enables one to determine the single ion anisotropy parameters
since magnetic ions are in the same surrounding
\cite{lattices1,lattices2,lattices3} as in the concentrated
magnet. We found the presence of a single-ion anisotropy,
evidenced by the obvious splitting of the ESR absorption spectrum,
and we determined the values of the main anisotropy constants.

\section{Experimental details and samples.}

Polycrystalline samples of
(Y$_{0.995}$Gd$_{0.005}$)$_2$Sn$_2$O$_7$ were prepared by heating
a stoichiometric mixture of Y$_2$O$_3$ (99.99\%), Gd$_2$O$_3$
(99.999\%) and SnO$_2$ (99.9\%) in air. The samples were heated
between 1400$^\circ$C and 1450$^\circ$C for several hours with
intermediate regrindings. The x-ray diffraction pattern obtained
at room temperature with a Bruker D8 diffractometer corresponds to
the pure phase with a  small amount ($<$1\%) of SnO$_2$. A
magnetic susceptibility curve was measured with a field of 80\,Oe
in the temperature range 20\,K-250\,K. Sample magnetization
corresponds to the Curie law with the effective moment of
7.9\,$\mu_{\mathrm{B}}$ per Gd$^{3+}$ ion, assuming the nominal Gd
concentration of 0.5\%.

The ESR study was performed at frequencies 18-100\,GHz using a set
of home-made transmission-type ESR spectrometers equipped with a
He-cooled cryomagnet. The ESR absorption spectra were recorded as
 field dependences  of the transmitted microwave signal.

\section{Experimental results and discussion.}

Electron spin resonance is a sensitive method for the
determination of the spin-Hamiltonian constants (see, e.g.,
\cite{abragam}). The microwave absorption power per spin is:

\begin{equation}\label{eqn:transint}
    P_{\mathrm{abs}}=\frac{\pi g^2\mu_{\mathrm{B}}^2\omega}{2\hbar}
    \sum_{E_j>E_i}\frac{e^{-E_i/T}-e^{-E_j/T}}{Z}|\langle i|
(\widehat{\bi
{S}}\cdot\bi{h})|j\rangle|^2\delta(\omega-\omega_{ji}),
\end{equation}
where $\bi{h}$ is the microwave field of the frequency $\omega$,
$E_i$, $E_j$ are the energies of the corresponding spin states,
$\omega_{ji}=(E_j-E_i)/\hbar$ and $Z=\sum\exp\{-E_n/T\}$.

The microwave field is polarized perpendicular to the external
magnetic field in our experiments. Thus, only dipolar transitions
with $\Delta S_{\mathrm{H}}=\pm1$ are observed ($S_{\mathrm{H}}$
is the spin projection onto the magnetic field direction). In the
absence of single-ion anisotropy, a single resonance line should
be observed at the resonance frequency
$\omega=g\mu_{\mathrm{B}}H/\hbar$. The presence of a single-ion
anisotropy results in a multi-component absorption spectrum.  The
amplitudes and the signs of the spin-Hamiltonian parameters may be
determined from the positions and the relative intensities of
these components.

\begin{figure}
  \centering
  \includegraphics{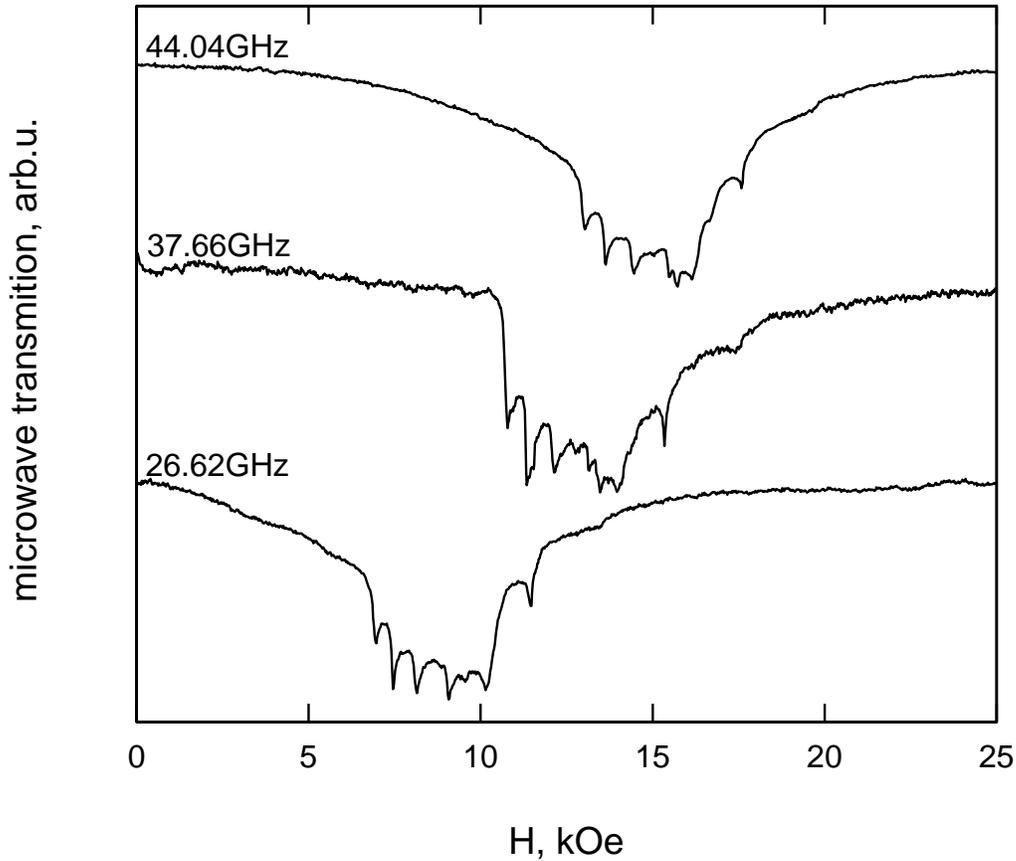}
  \caption{ESR absorption spectra obtained in \ygdsn{} at 4.5\,K with
   different frequencies.}
\label{fig:lines-f}
\end{figure}

\begin{figure}
  \centering
  \includegraphics{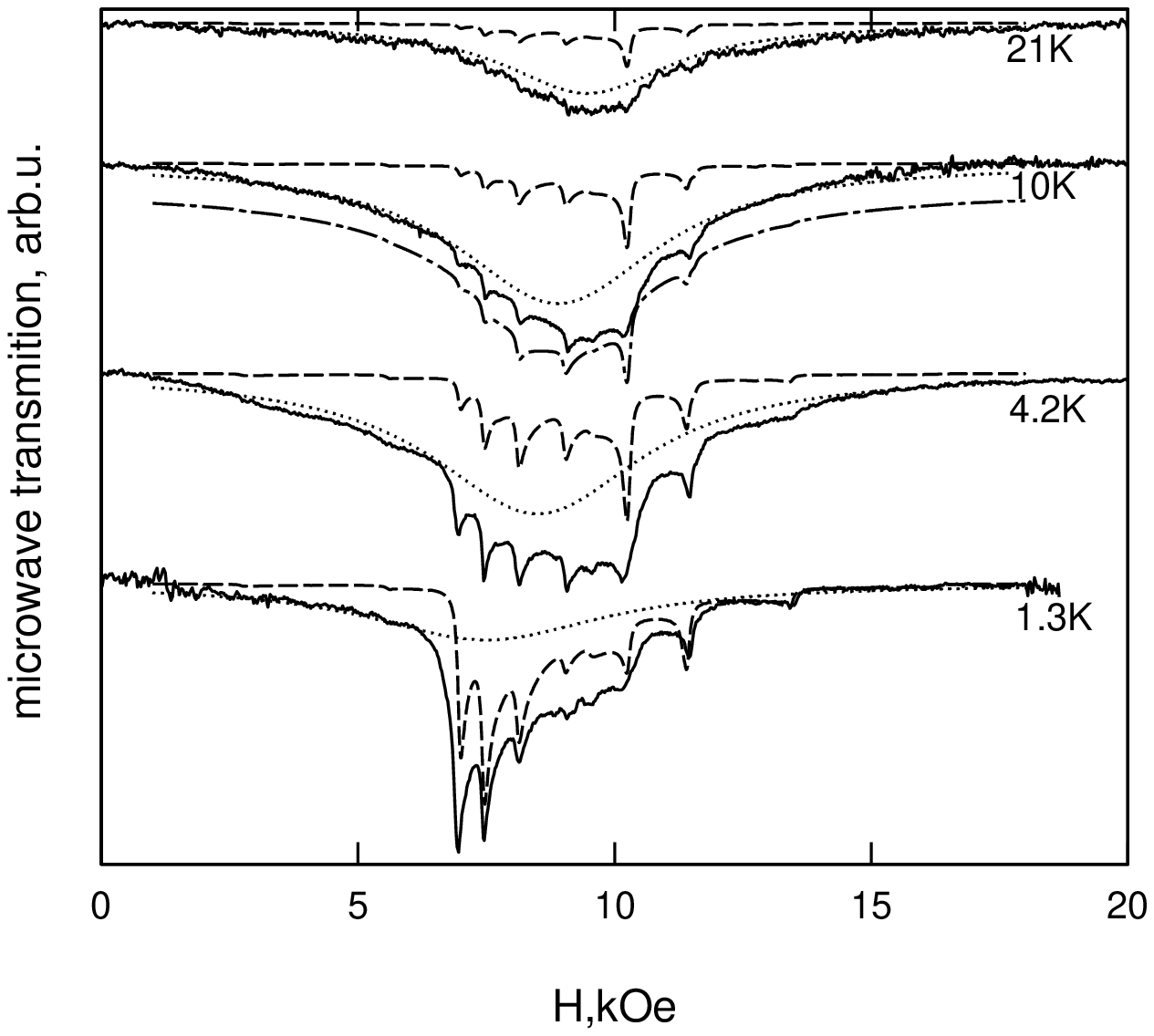}
  \caption{ESR absorption spectra obtained in \ygdsn{} with f=26.62GHz at
  different temperatures (solid lines); calculated ESR absorption spectra
  (dashed lines); additional broad absorption signal (dotted lines);
  at 10\,K, sum of the calculated ESR absorption and of an additional broad
  absorption line (dash-dotted line, shifted for clarity).}
\label{fig:lines-t}
\end{figure}

ESR absorption spectra measured at different microwave frequencies
and at different temperatures are shown in Figures
\ref{fig:lines-f}, \ref{fig:lines-t}. The absorption spectrum is
clearly multi-component, which indicates the presence of crystal
field splitting of the single-ion energy levels.

For    powder samples the  absorption is averaged over all
possible orientations of the crystallographic  axes. The point
symmetry of the Gd ion surroundings is $D_{3d}$, and the main term
in the anisotropy energy is the second order axial term with the
anisotropy axis along the $\langle111\rangle$ direction.
Measurements on the single crystals of the pyrochlore titanate
have shown that this term is the most important \cite{glazkov}. As
a first step of the present analysis, we will consider a
simplified model restricted to this term only:

\begin{equation}\label{eqn:ham-simple}
    {\cal H}=g\mu_{\mathrm{B}}\bi{H}\cdot\widehat{\bi{S}}+DS_z^2.
\end{equation}
Treating the anisotropy term in the first order of perturbation
theory, for the magnetic field applied at the angle $\theta$ with
respect to the anisotropy axis $z$, we obtain for the energies, up
to a constant term:

\begin{equation}\label{eqn:energy-perturb}
    E(m)=(g\mu_{\mathrm{B}}H)m+\frac{D}{2}(3\cos^2\theta-1)m^2,
\end{equation}
where $m$ is the spin projection on the direction of the
magnetic field. The resonance field corresponding to the
$|m\rangle\leftrightarrow|m+1\rangle$ dipolar transition is

\begin{equation}\label{eqn:H-perturb}
    H_m(\theta)=H_0-\frac{D}{2g\mu_{\mathrm{B}}}(3\cos^2\theta-1)(2m+1),
\end{equation}
where $H_0=\hbar\omega/(g\mu_{\mathrm{B}})$ is a free spin
resonance field. When averaging over the orientation, grains with
the anisotropy axis perpendicular to the field direction enter
with the higher weight. For the case of a narrow resonance line,
this averaging yields:

\begin{equation}\label{eqn:intensity-perturb}
    I_m(H)\propto \frac{1}{\sqrt{\frac{H_0-H}{\Delta H_m}+1}},
\end{equation}
where $\Delta H_m=D(2m+1)/(2g\mu_{\mathrm{B}})$, $m\neq-1/2$. The
magnetic fields, at which the resonance absorption is observed,
are limited by the values  $H_0-2\Delta H_m$ and $H_0+\Delta H_m$,
corresponding to grains with $\theta=0$ and $\pi/2$, respectively.
Thus, the averaged ESR absorption line corresponding to a given
$|m\rangle\leftrightarrow|m+1\rangle$ transition should
demonstrate sharp edges both at the left and at the right of the
paramagnetic resonance position. At the  end corresponding to
grains with $\theta=0$ it should have a step-like edge, while at
the other end ($\theta=\pi/2$ grains), it should demonstrate a
sharp increase of absorption corresponding to the square root
divergency in \eref{eqn:intensity-perturb}. Whether this increase
in absorption is to the left or to  the right from the free spin
resonance position $H_0$ depends, for a given $m$, on the sign of
the $D$ constant only.

We have studied the ESR absorption at different temperatures (see
\Fref{fig:lines-t}). At low temperatures, the contributions to the
absorption arising from the states with a large negative $m$
($-7/2$ and $-5/2$) dominate. In \Fref{fig:lines-t} one can
clearly see that an intense absorption at low temperatures occurs
on the left edge of the ESR spectra, while on the right edge there
are several step-like features of smaller amplitude. This
distribution of the intensities corresponds to a positive sign for
the $D$ constant, i.e. to an easy-plane anisotropy.

The above considerations suggest a way to identify the ESR absorption
spectrum features corresponding to $\theta=0,~\pi/2$. Intense
slightly asymmetric peaks observed at the left of the paramagnetic
resonance position at low temperatures and peaks of similar
shape that become more intense on heating correspond to
$\theta=\pi/2$. Step-like changes of the absorption observed at
the left and at the right of the paramagnetic resonance
position, and intense features observed at low temperatures at the
right of the paramagnetic resonance position correspond to
grains with $\theta=0$.

\begin{figure}
  \centering
  \includegraphics{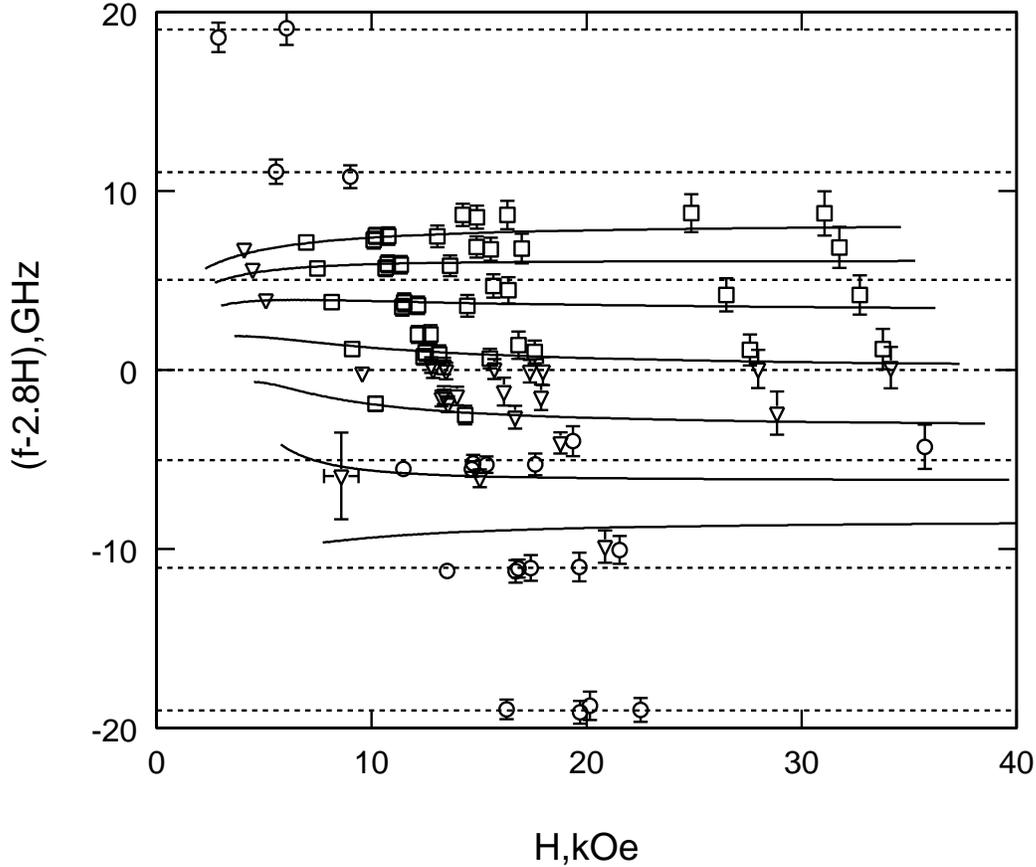}
  \caption{Frequency-field dependences for the different lines
  of the observed ESR absorption spectra (symbols) and calculated dependences
  (lines). A linear quantity $f=g\mu_{\mathrm{B}} H$, with $g=$2.0, has been
  subtracted. $\theta=0$ features (circles); $\theta=\pi/2$ features
  (squares);
  features of the ESR absorption which were hard to identify (triangles);
  calculated frequency-field dependences for $\theta=\pi/2$ (solid lines);
  calculated frequency-field dependences for $\theta=0$ (dashed lines).}
\label{fig:f(h)}
\end{figure}

The frequency-field diagram of the observed transitions is shown
in \Fref{fig:f(h)}. Note that the $\theta=0$ components do not
form equidistant lines. This indicates the presence of higher
order contributions to the anisotropy energy. The general form of
the spin-Hamiltonian for the S=7/2 Gd$^{3+}$ ion in the
surroundings of $D_{3d}$ symmetry is:
\begin{equation}
    {\cal H}=g\mu_{\mathrm{B}}\bi{H}\cdot\widehat{\bi{S}}
+\BStev{0}{2}+\BStev{0}{4}+\BStev{3}{4}+\BStev{0}{6}
+\BStev{3}{6}+\BStev{6}{6}\label{eqn:ham-generic}
\end{equation}
where the Stevens operators \cite{abragam} $\Stev{i}{j}$ are
functions of the components of the total angular momentum $S=7/2$,
with ${\mathbf z}\parallel\langle111\rangle$ and ${\mathbf
x}\parallel\langle 11\overline{2}\rangle$. In the notations of
\eref{eqn:ham-simple}: $D=3B_2^0$.

We will consider only the second and fourth order axial terms, and
we will suppose $g=2.0$ in the further analysis. The $B^0_2$ and
$B^0_4$ constants can be found from the positions of the
well-defined $\theta=0$ components corresponding to the
$|-7/2\rangle\leftrightarrow|-5/2\rangle$  and
$|-5/2\rangle\leftrightarrow|-3/2\rangle$ transitions (two
lowermost lines in \Fref{fig:f(h)}). Deviations of these
transitions frequencies from the paramagnetic resonance position
are $19.0\pm0.3$\,GHz and $11.2\pm 0.3$\,GHz correspondingly. This
gives the following values for the spin-Hamiltonian constants:
$B^0_2=(47\pm1)$mK and $B^0_4=(0.05\pm0.02)$mK. The amplitude of
the second order anisotropy constant ($D=3B^0_2$=140\,mK) is 35\%
smaller than the corresponding value in \gdti{} ($D=223$\,mK). The
frequency-field dependences for $\theta=0,~\pi/2$ calculated for
these parameter values are shown in  \Fref{fig:f(h)}.
Additionally, we have modeled the ESR absorption line using
\eref{eqn:transint}. The simulation was performed by exact
numerical diagonalization of the Hamiltonian matrix combined with
the averaging over the orientations of the grains crystallographic
axes. The results are presented in \Fref{fig:lines-t}. The
calculated frequency-field dependences are in a good agreement
with the observed ones. The simulated absorption spectra also
demonstrate good agreement with the experimental data
 --- they correctly reproduce the details of the line shape as well as
the tendencies of the line shape change with the temperature.

Note that the correspondence between the simulated and observed
absorption spectra can be improved by adding a broad absorption
line located near the paramagnetic resonance position (see
\Fref{fig:lines-t}). This additional contribution cannot be
attributed to isolated Gd$^{3+}$ spins in the crystal field. The
intensity of this additional absorption component increases with
increasing temperature at low temperatures: the integral intensity
of this broad line at 4.2K is twice as large as its integral
intensity at 1.3K. The ratio of the intensities of the suggested
broad absorption line and of the crystal-field split absorption is
also temperature dependent:  at 1.3K the most of  the integral
intensity is due to the crystal-field split absorption lines,
while at 20K the broad absorption line dominates. A possible
origin of this broad absorption line is the formation of
antiferromagnetically coupled pairs of Gd ions. The ground state
of a pair is a nonmagnetic singlet, the total splitting between
the S=0 ground state and the highest energy state (S=7) is about
$2JS_{Gd}^2\sim7$K. The dominating intensity of the broad line at
20K suggests a high concentration of these pairs, which indicates
a tendency toward formation of  Gd clusters in \pyr{Y}{Sn}{}
matrix. However, the characteristic multi-component signal
corresponding to the isolated S=7/2 Gd$^{3+}$ ions can be
distinguished and analyzed at low temperatures. The presence of
exchange coupled pairs has also been observed by $^{170}$Yb
M\"ossbauer spectroscopy in \pyr{Y}{Ti}{} doped with Yb
\cite{hodg}.

\section{Conclusions.}

We report on the ESR in powder samples of \ygdsn. The observed
splitting of the ESR absorption line allows to determine the
values of the single-ion anisotropy constant for the Gd ion. We
find the value of the main anisotropy constant is about 35\%
smaller than that in titanate \gdti{}. Possibly, this change of
the anisotropy energy is responsible for the strong difference of
the ordered state formed in \gdti{} and \gdsn{} at low
temperatures.

\ack
 Work was supported by the grant of  Russian Foundation for
Basic Research No.04-02-17294.

\end{document}